\documentclass[hyper]{JHEP3}

\usepackage{epsfig}
\usepackage{multicol}
\usepackage{bbm}
\usepackage{amsmath}
\usepackage{cite}
\usepackage{lscape}
\usepackage{multirow}
\usepackage{amscd} 
\usepackage{slashed}
\usepackage{enumerate}

\newcommand{\rinv}{$R^{-1}$~}
\newcommand{\lambdar}{$\Lambda R $~}

\newcommand{\gev}{\mbox{ GeV}}
\newcommand{\tev}{\mbox{ TeV}}
\newcommand{\cf}{{\it c.f. }}

\allowdisplaybreaks

\title{Constraints on models with universal extra dimensions from dilepton searches at the LHC}

\author{
Lisa Edelh\"auser${}^\ddagger$, Thomas Flacke${}^\dagger$, Michael Kr\"amer${}^\ddagger$ \\
${}^\ddagger$Institute for Theoretical Particle Physics and Cosmology, RWTH Aachen University,
Aachen, Germany\\
${}^\dagger$Department of Physics, Korea Advanced Institute of Science and Technology,
335 Gwahak-ro, Yuseong-gu, Daejeon 305-701, Korea\\
E-mail: ledelhaeuser@physik.rwth-aachen.de, flacke@kaist.ac.kr, mkraemer@physik.rwth-aachen.de}

\abstract{
Models with universal extra dimensions predict that each Standard Model particle is accompanied by a tower of Kaluza-Klein resonances. 
Canonical searches for the production and cascade decays of first Kaluza-Klein modes through missing transverse momentum signatures 
suffer in general from low detection efficiencies because of the rather compressed Kaluza-Klein particle mass spectrum. 
Here, we instead analyze signatures from the production of second Kaluza-Klein states which can decay into Standard Model particles and thus do not 
result in any missing transverse momentum. Such signatures provide a strong sensitivity, and are of particular interest as they 
would allow for a clear distinction between extra dimensional models and other new physics such as minimal supersymmetric models. 
We constrain the production of second Kaluza-Klein particles from recent LHC searches for dilepton resonances, and place limits on 
the compactification scale  $R^{-1} \gtrsim 715$\,GeV and on the masses of the second Kaluza-Klein particles  $M_{{\rm KK}^{(2)}} \gtrsim 1.4$\,TeV in the minimal UED model.}

\preprint{TTK-13-05 
          } 

\begin{document}
\section{Introduction}\label{sec:intro}

Models with universal extra dimensions (UED) \cite{Appelquist:2000nn, Hooper:2007qk} represent the simplest extra-di\-men\-sio\-nal extension of the Standard Model with a dark matter candidate and potentially rich LHC phenomenology.\footnote{For earlier proposals of TeV scale extra dimensions \cf Ref.~\cite{Antoniadis:1990ew}.} All fields are assumed to propagate in a flat space-time $\mathcal{M}_4\times X$, where $X$ is a compact space. Due to the compactification, the momentum along the extra-dimension(s) is discretized. In the four-dimensional effective theory, this leads to a tower of Kaluza-Klein (KK) resonances in the particle spectrum. Each  Standard Model particle, which is identified with the lowest-lying (``zero'') mode, is therefore accompanied by a KK tower of heavier partners with the same quantum numbers and in particular with the same spin. 

In this article, we focus on the extra dimension $X=S^1/Z_2$ with a compactification radius $R$, which represents the simplest choice allowing for a standard-model-like zero-mode spectrum.\footnote{For other compactifications \cf Refs.\cite{Appelquist:2000nn,6DUED,Cacciapaglia:2012dy}.} The presence of the boundaries of the $S^1/Z_2$ fundamental domain breaks translational invariance along the extra dimension and thereby leads to the violation of Kaluza-Klein-number (``KK-number'')  in interactions amongst the KK-excitations of the 4D effective theory. Nevertheless, as long as boundary conditions and interactions are chosen symmetrically at both boundaries, the model still respects a $Z_2$ symmetry (``KK-parity''). KK-parity guarantees the stability of the lightest KK-excitation (LKP) which provides the model with a viable dark matter candidate \cite{UEDDM,Belanger:2010yx}.

As an extra-dimensional theory, the 5D UED model is inherently non-renormalizable and must therefore be considered as an effective theory with a cutoff $\Lambda$. At the scale $\Lambda$ higher-dimensional operators in the bulk and on the boundaries parameterize our ignorance about the UV completion of the model. Note that unitarity \cite{SekharChivukula:2001hz}, naive dimensional analysis \cite{Flacke:2005hb}, and the stability of the Higgs vacuum expectation value \cite{Blennow:2011tb} imply a low cutoff of $\Lambda R\sim \mathcal{O}(10)$, where $R^{-1}$ is the compactification scale of the extra dimension. In this paper, we focus on the most commonly studied minimal UED model (MUED) \cite{Cheng:2002iz}, in which only the 5D extensions of the Standard Model operators are present at the cutoff scale $\Lambda$, while boundary operators and other higher-dimensional bulk operators are assumed to be vanishing at $\Lambda$.\footnote{For studies of non-minimal bulk- or boundary operators, \cf Refs.~\cite{nUED}.} Under this assumption, the only free parameters of the MUED model are the com\-pac\-ti\-fi\-ca\-tion scale $R^{-1}$ and the cutoff scale $\Lambda R$, while all other parameters are fixed by the matching of the zero mode masses and interactions with the Standard Model particles. In the absence of non-minimal operators, the interactions between KK modes are KK-number-preserving and equal to the Standard Model couplings, up to loop corrections. KK-number violating operators are induced at one-loop level \cite{Georgi:2000ks,Cheng:2002iz}, while KK-parity remains an exact symmetry. 

The phenomenology of the 5D MUED model has been studied extensively. The currently strongest lower limit on the compactification scale arises from electroweak precision tests \cite{EWbounds,Belyaev:2012ai} which yield a bound of $R^{-1}\gtrsim ~750 \gev$ if the recently discovered boson with a mass of $125 \gev$  \cite{Higgs} is a UED Higgs boson \cite{UEDHiggs}. Bounds from FCNCs imply $R^{-1}\gtrsim 650 \gev$ \cite{UEDFlavor}.  An upper bound on the compactification scale follows from the relic density of the dark matter candidate, because a too heavy dark matter candidate implies early decoupling which leads to over-closure of the universe. However, as has been shown in Ref.\cite{Belanger:2010yx}, the upper bound of $R^{-1}\lesssim 1.6 \tev$ sensitively depends on the KK mass spectrum because both, co-annihilation processes with first KK mode particles, and resonant annihilation through second KK mode particles may play a crucial role in determining the relic density.

With the mass scale of the first KK modes being constrained to lie around $1 \tev$ from pre-LHC bounds, the MUED model  provides an interesting candidate for direct LHC searches. Considering only the zeroth and first KK level, UED predicts a partner particle to every Standard Model particle with opposite KK parity and therefore signatures qualitatively similar to those of the minimal supersymmetric model (MSSM)~\cite{mssm}. Differences between MUED and MSSM signatures arise from the difference in the spin of the KK and MSSM particles, the higher KK production cross section, a rather compressed first KK mode mass spectrum, and a differing Higgs partner sector. Focussing on MSSM-like missing $\slashed{E}_T$ signatures, several studies of the discovery- and exclusion-reach of the MUED model have been performed \cite{UEDsearches,Datta:2005zs,Datta:2010us}. For the tri-lepton channel, Ref.~\cite{Belyaev:2012ai} projects an exclusion reach of $R^{-1}\gtrsim 1.2 \tev$ for $20\,{\rm fb}^{-1}$ of LHC data at 8 TeV. A more generic analysis of LHC $\slashed{E}_T$ constraints on models with universal extra dimensions has been presented in Ref.~\cite{Cacciapaglia:2013wha}. 
Searches for production and cascade decays of first KK mode particles, however, suffer in general from low detection efficiencies. The rather compressed UED mass spectrum implies that the Standard Model particles resulting from the cascade are soft, while the resulting LKP's at the end of the decay chain typically have opposite $p_T$ so that events generically have low total $p_{T,{\rm miss}}$. Such signatures are therefore hard to distinguish from Standard Model background with canonical MSSM search cuts, and specific UED-tailored cuts are required to improve the sensitivity of $\slashed{E}_T$ searches for KK mode particles.

In this article we instead focus on signatures arising from second KK mode interactions \cite{Cacciapaglia:2012dy,Datta:2005zs,Chang:2012wp}. These signatures are of particular interest because, if detected, they allow for a clear distinction between UED and the MSSM. In contrast to first KK modes, second KK modes have even KK parity and can therefore decay into Standard Model particles without leading to missing energy in the event. Here, we focus on the channel $pp\rightarrow A^{(2)}/Z^{(2)}+X\rightarrow l \bar{l}+X'$ and constrain the model by comparing the predicted MUED rates to the results of the $Z'$ search in the dilepton channel by CMS \cite{cmsdilepton}.\footnote{Searches for second KK mode $W^{(2)}$ and gluons $g^{(2)}\rightarrow t \overline{t}$ have been considered in Refs.~\cite{ Flacke:2012ke,Chang:2012wp} .} To calculate the event rates, we extended the implementation of the MUED model \cite{Christensen:2009jx}  in Feynrules \cite{Christensen:2008py} and included the second KK modes and the corresponding interactions. We also included the KK-number violating interactions for the second KK gauge boson modes according to \cite{Cheng:2002iz,Datta:2010us}, and took into account the one-loop radiative corrections for the KK mass spectrum according to \cite{Cheng:2002iz}. 

This article is structured as follows. In the next section, we review the main features of MUED second KK mode particles, their masses and interactions, and discuss the implications for our search. In particular, we also discuss the different production channels for $Z^{(2)}$ and $A^{(2)}$ modes and show that the main contribution arises from strong production of second KK mode quarks and gluons which cascade decay into them. In Sec.~\ref{sec:results}, we present the detailed study of the  $pp\rightarrow A^{(2)}/Z^{(2)}+X\rightarrow l \bar{l}+X'$ channel and the resulting bounds on the MUED parameter space. We summarize and conclude in Sec.\ref{sec:summary}.

\section{The second KK level of the MUED model}\label{sec:model}
In our implementation, we follow Ref.\cite{Datta:2010us}, where the  expressions for the one-loop masses and a list of Feynman rules can be found. In this section we wish to highlight some features of the mass spectrum and interactions which are relevant for our analysis.

The masses  of  the $n^{\rm th}$ KK partner of Standard Model gauge bosons $\mathcal{G}$ and fermions $\psi$ are determined from \cite{Cheng:2002iz} 
\begin{eqnarray}
m^2_{\mathcal{G}^{(n)}}&=&\left(\frac{n}{R}\right)^2+a_\mathcal{G} \frac{\zeta(3)}{16 \pi^4}\frac{1}{R^2}+\frac{b_{\mathcal{G}}}{16 \pi^2}\ln \left(\frac{\Lambda^2}{\mu^2}\right)\frac{n^2}{R^2},\label{eq:mass}\\
m_{\psi^{(n)}}&=&\left(\frac{n}{R}\right)+\frac{b_{\psi}}{16 \pi^2}\ln \left(\frac{\Lambda^2}{\mu^2}\right)\frac{n}{R},\label{eq:mass2}
\end{eqnarray}
where $R$ denotes the compactification radius, $\Lambda$ denotes the cutoff of the theory which we treat as an input parameter, and $\mu$ denotes the renormalization scale. Contributions from electroweak symmetry breaking are subdominant, such that the mass eigenstates are to a very good approximation  given by the gauge eigenstates of the gauge bosons $G^{(n)}, W^{(n)}, B^{(n)}$ and the Dirac fermions $Q^{(n)},U^{(n)},D^{(n)},L^{(n)},E^{(n)}$.\footnote{Details of the mass diagonalization for the neutral electroweak gauge bosons and third family quarks are discussed in Ref.~\cite{Cheng:2002iz} and taken into account in our implementation.} The coefficients $a_{\mathcal{G}} $ and $b_{\mathcal{G},\psi} $ are collected in Table \ref{table:masses}. The first contribution in Eqs.~(\ref{eq:mass}, \ref{eq:mass2}) arises from the compactification at tree-level, while  the second contribution (absent for fermions at one-loop level) is a KK mode-independent one-loop ``bulk'' correction. The third term is due to ``brane-localized'' corrections related to the orbifolding, which introduce the dependence on $\Lambda/\mu$, or, when choosing the renormalization scale at the second KK mode mass-scale $2R^{-1}$, the dependence on the dimensionless parameter $\Lambda R$. As can be seen, the overall mass scale of the second KK modes is set by $2 R^{-1}$ while the relative mass splitting between different particles at the second KK level is controlled by $\ln ((\Lambda R)^2/4)$. The heaviest second KK mode particle is the partner of the gluon and the lightest particle is, to a good approximation, the partner of the $U(1)_Y$ gauge boson.

\begin{table}\centering
\renewcommand{\arraystretch}{1.5}
\begin{tabular}{|c|c|c|c|c|c|c|c|c|}
\hline
& $G^{(n)}$ & $W^{(n)}$ & $B^{(n)}$ & $Q^{(n)}$ & $U^{(n)}$ & $D^{(n)}$ & $L^{(n)}$ & $E^{(n)}$ \\\hline
$a$ & \scriptsize $-\frac{3}{2}g_3^2$ &\scriptsize $-\frac{5}{2}g_2^2$ &\scriptsize $-\frac{39}{2}g_1^2$ &\scriptsize 0 & 0 & \scriptsize0 &\scriptsize 0 &\scriptsize 0 \\\hline 
$b$ &\scriptsize$\frac{23}{2}g_3^2$ &\scriptsize $\frac{15}{2}g_2^2$  & \scriptsize$-\frac{1}{6}g_1^2$ & \scriptsize $3 g_3^2 +\frac{27}{16}g_2^2+\frac{1}{16}g_1^2$ &\scriptsize $3 g_3^2 +g_1^2$  &\scriptsize $3 g_3^2 +\frac{1}{4}g_1^2$  &\scriptsize $\frac{27}{16}g_2^2+\frac{9}{16}g_1^2$ & \scriptsize$\frac{9}{4}g_1^2$ \\\hline
\end{tabular}
\caption{One-loop induced mass correction parameters in MUED \cite{Cheng:2002iz}. $g_{3,2,1}$ denote the $SU(3)$, $SU(2)_L$ and $U(1)_Y$ gauge couplings.}
\label{table:masses}
\end{table}

KK number conserving couplings, which are present already at tree level, are of the strength of the corresponding Standard Model couplings (in the gauge eigenbasis). KK number violating (but KK parity preserving) couplings are induced only at the one-loop level and can be expressed in terms of the boundary contribution coefficients as \cite{Cheng:2002iz} 
\begin{equation}
\mathcal{L}\supset -ig_\mathcal{G}\left[\frac{b_{\mathcal{G},2}-2 b_{\psi,2}}{\sqrt{2}\, 16 \pi^2}\ln \left(\frac{\Lambda^2}{\mu^2}\right) \right]\overline{\psi}^{(0)} \gamma^\mu T^a P_+\psi^{(0)} \mathcal{G}^{a(2)}_\mu.\label{eq:coupl}
\end{equation}
Here $T^a$ is the generator associated to the gauge group of $\mathcal{G}$ and $P_+$ is $ P_R$ $(P_L)$ for fermions with a right- (left-) handed zero mode. Choosing again $\mu=2 R^{-1}$ as the renormalization scale, the KK-number violating interaction in Eq.~(\ref{eq:coupl}) is proportional to $\ln((\Lambda R)^2/4)$.  In contrast, KK-number violating triple gauge boson vertices like $G^{(0)}G^{(0)}G^{(2)}$ and KK-number violating interactions like $\mathcal{G}^{(0)}\overline{\psi}^{(0)}\psi^{(2)}$ with one fermion zero and one fermion two-mode are not induced \cite{Cheng:2002iz}.

\begin{figure}[htp]\centering
\includegraphics[height=0.17\textwidth]{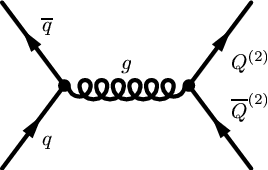}\hspace{5ex}
\includegraphics[height=0.17\textwidth]{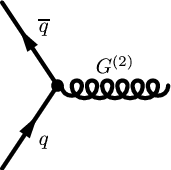}\hspace{5ex}
\includegraphics[height=0.17\textwidth]{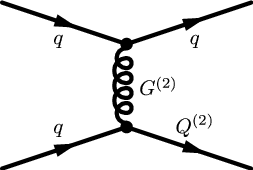}
\caption{Examples of Feynman diagrams for KK-number preserving, KK-number violating direct, and KK-number violating associated production (from left to right). $Q^{(2)}$ and $G^{(2)}$ represent a second KK singlet or doublet quark and a second KK gluon mode, respectively.}
\label{fig:channels}
\end{figure}

This pattern of masses and couplings has an important impact on the production and the decays of second mode KK particles. For the production of strongly interacting second KK modes, three main production mechanisms compete (see Figure \ref{fig:channels}):
\begin{enumerate}[(A)]
\item \label{prod1} \emph{KK-number preserving production:}
The non-resonant  KK number-preserving production of two second KK modes through quark-quark, quark-gluon, or gluon-gluon parton scattering; 
this process suffers from phase-space suppression because two heavy modes are produced.
\item \label{prod2} \emph{KK-number violating direct production:}
Resonant production of a second KK gluon is one-loop suppressed (because of the KK number violating coupling), and reduced because of the parton distribution functions, as a $q\bar{q}$ is required in the initial state.
\item \label{prod3} \emph{KK-number violating associated production:}
A second KK mode quark or gluon can be produced in association with another zero mode. These non-resonant processes again are one-loop suppressed because of the KK-number violating coupling.
\end{enumerate}

The production of second KK leptons and electroweak gauge bosons dominantly occurs due to cascade decays of the strongly produced second KK mode(s), as well as -- in the case of gauge boson production -- direct $s$-channel production analogous to type (\ref{prod2}) production. 

\begin{figure}[htp]\centering
\includegraphics[width=1\textwidth]{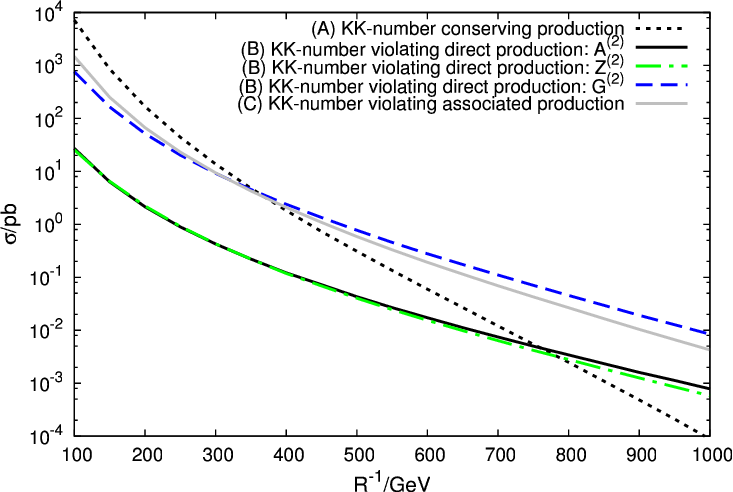}
\caption{8 TeV LHC production cross sections for second KK quark and gluon modes from KK-number preserving, KK-number violating direct, and KK-number violating associated production with  \lambdar= 20. For comparison we also show the cross section for the direct production of $A^{(2)}$ and $Z^{(2)}$. The CTEQ6L paton distribution functions \cite{Pumplin:2002vw} have been adopted and the renormalization and factorization scales have been set to $\mu =2R^{-1}$. }
\label{fig:producedcross}
\end{figure}

Figure~\ref{fig:producedcross} shows the LHC 8 TeV production cross sections for second KK modes through the different production mechanisms as a function of $R^{-1}$ with the cutoff scale set to $\Lambda R=20$. At small $R^{-1}$ second KK mode quarks and gluons are mostly produced by KK-number preserving interactions (\ref{prod1}). At large $R^{-1}$, strongly coupled second KK modes are dominantly produced via KK-mode violating direct (\ref{prod2}) and associated (\ref{prod3}) channels. The direct production of $A^{(2)}$ and $Z^{(2)}$ is suppressed because it arises from $U(1)_Y$ and $SU(2)$ gauge interactions.  For smaller $\Lambda R$, the KK-number preserving channels are enhanced, because the second KK mode masses are reduced (\cf Eq.~(\ref{eq:mass})). The KK-number violating channels are reduced because of the smaller KK-number violating coupling Eq.~(\ref{eq:coupl}).

\begin{figure}[htp]\centering
\includegraphics[width=0.3\textwidth]{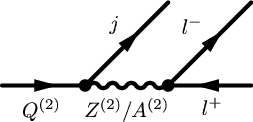}
\caption{Generic decay of a second KK mode quark into a jet and two leptons.}
\label{fig:decay}
\end{figure}

The decay of second KK modes can proceed through KK-number conserving interactions into one second  and one zero mode or into two first KK modes, as well as through KK-number violating decays into two zero modes, see Ref.~\cite{Datta:2005zs} for a detailed discussion. The decay patterns and features most relevant to our study are:
\begin{itemize}
\item second KK mode gluons $G^{(2)}$ have a considerable branching fraction ($\sim  40\%$) into second KK mode $SU(2)$ doublet and singlet quarks $Q^{(2)}$ and $U^{(2)} / D^{(2)}$,
\item second KK mode $SU(2)$ singlet quarks have a branching fraction of $\sim  50\%$ into $A^{(2)}$,
\item second KK mode $SU(2)$ doublet quarks have a branching fraction of $\sim  20\%$ into $Z^{(2)}$.
\end{itemize}
The decays of second KK quarks into $A^{(2)}$ or $Z^{(2)}$ may then lead to dilepton resonances as shown in Figure \ref{fig:decay}. The branching fractions of the $A^{(2)}$ and $Z^{(2)}$ into $e^+e^-$ are $\sim 0.5 \%$ with the same branching fractions into $\mu^+\mu^-$. 

\section{Signal simulation and dilepton resonance limits}\label{sec:results}
\subsection{CMS limits in dilepton searches}

The CMS  exclusion limit on new resonances in the dilepton spectrum \cite{cmsdilepton} is expressed in terms of the parameter 
\begin{eqnarray}
\mathcal{R}=\frac{\sigma_{Z'}^{\mbox{ \tiny w/o C}}}{\sigma_{Z}^{\mbox{\tiny w/o C}}}=\frac{\sigma \left(pp\rightarrow Z'+X\rightarrow ll+X\right)}{\sigma \left(pp\rightarrow Z+X\rightarrow ll+X\right)}.	
\label{eq:rfactor}
\end{eqnarray}
The cross section $\sigma_{Z'/Z}^{\mbox{ \tiny w/o C}}$  is  the theoretical cross section without any cuts. It is obtained from data by using a Monte-Carlo acceptance factor. We have checked that the acceptance factor for the various contributions 
to the MUED signal (direct, associated,  and KK-number conserving) differs by less then 5\% from the acceptance factor for a standard sequential $Z'$. In the following simulation, we therefore determine the signal cross sections {\it without} cuts, since the acceptance is already incorporated in the exclusion limits through the parameter $\mathcal{R}$.  We use the computer program {\tt FEWZ\,3}~\cite{Li:2012wn,Gavin:2010az} for the calculation of the NNLO $Z$ cross section $ \sigma \left(pp\rightarrow Z+X\rightarrow ll+X\right)$  needed in the denominator of the $\mathcal{R}$ parameter of Eq.~(\ref{eq:rfactor}).
\subsection{Simulation and results}

As discussed in the previous sections, we have extended the implementation of MUED in Feynrules~\cite{Christensen:2008py,Christensen:2009jx}  by including the second KK mode particles with masses and interactions as outlined in Refs.~\cite{Cheng:2002iz,Datta:2010us}.  We use {\tt Madgraph\,5}~\cite{Alwall:2011uj} with the CTEQ6L parton distribution functions \cite{Pumplin:2002vw} for the simulation of the various processes at leading-order. The theoretical uncertainty on the production cross section is estimated by varying the renormalization and factorization scales by a factor of two around the central value $\mu = 2R^{-1}$. The exclusions presented below are conservatively based on the lower value of the cross section prediction. 
We calculate the masses and decay widths of the particles according to Ref.~\cite{Cheng:2002iz}.

MUED predicts two resonances in the $ l l$ invariant mass spectrum, coming from the (lighter) $A^{(2)}$ and the (heavier) $Z^{(2)}$ decays. The $A^{(2)}$ resonance in the $ll X$ final state is higher by a factor $\approx 1.6$, which is mostly due to the larger branching ratio of $U^{(2)} / D^{(2)}$ decaying into $A^{(2)}$ as compared to the branching ratio of $Q^{(2)}$ decaying into $Z^{(2)}$. Theoretically these two resonances are separable as their widths are smaller than their mass difference.\footnote{This also excludes (negative) interference effects, which however in UED have been shown to be highly suppressed anyway (\cf Ref.~\cite{Rizzo:2012rb}).}  However, the bin size used in the CMS analysis \cite{cmsdilepton} for the resonance scale relevant here ($\approx 1 \tev$) is larger than the mass difference $m_{Z^{(2)}}-m_{A^{(2)}}$. To combine the effect of two nearby peaks is non-trivial  because CMS uses shape variables in their analysis. Therefore  we conservatively only include leptons from an $A^{(2)}$ resonance as our signal cross-section $\sigma^{\mbox{\tiny w/o C}}_{Z'}$.

\begin{figure}\centering
\includegraphics[width=1\textwidth]{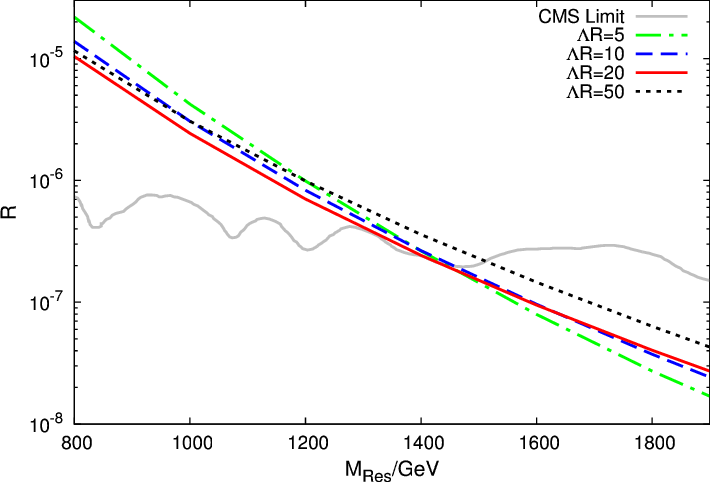}
\caption{The ratio $\mathcal{R}$ defined in Eq.~(\protect\ref{eq:rfactor}) for benchmark points \lambdar= 5, 10, 20, 50, as a function of the 
resonance mass $M_{\rm Res}\equiv m_{A^{(2)}}$.}
\label{fig:excl1}
\end{figure}

\begin{table}\centering
\begin{tabular}{|c|c|c|c|c|}
\hline
$\Lambda R$ & 5 & 10 & 20 & 50 \\\hline
$R^{-1}/[\gev] $& 720 & 730 & 715 &  755 \\ \hline
\end{tabular}
\caption{95\% C.L.\ lower bounds on $R^{-1}$ for different cutoff parameters $\Lambda R$.}
\label{tab:Rinvbounds}
\end{table}

Figure \ref{fig:excl1} shows our results for the cross-section ratio $\mathcal{R}$ defined in Eq.~(\ref{eq:rfactor}) as a function of the resonance mass $M_{\rm Res}\equiv m_{A^{(2)}}$ for MUED with $\Lambda R= \{5,10,20,50\}$, together with the CMS bound. The corresponding bounds on the compactification scale $R^{-1}$ are listed in Table~\ref{tab:Rinvbounds}.\footnote{For \lambdar=20, the lower bound on \rinv is 715 GeV, with a small window of $655\gev < R^{-1} < 660 \gev$ being only marginally excluded.} As can be seen, the bounds show a mild, non-monotonic  dependence on the cutoff scale $\Lambda R$, which arises from the relative weight of different production channels. For low $\Lambda R$,  KK-number conserving interactions are enhanced because the strongly coupling second KK resonances are lighter. For example, for  $\Lambda R=5$ they yield the dominant contribution to $A^{(2)}$ production up to a resonance mass of $M_{\rm Res}\approx 1.5$ TeV. For high $\Lambda R$, the KK-number violating interactions are enhanced. With $\Lambda R=50$, they dominate over the KK-number conserving production of $A^{(2)}$ modes already at a resonance mass of approximately $1$\,TeV.

\section{Summary and conclusion}\label{sec:summary}
Models with universal extra dimensions (UED) represent the simplest extra-di\-men\-sio\-nal extension of the Standard Model. They predict  a tower of Kaluza-Klein resonances, the lightest of which is weakly interacting and stable and thus a viable dark matter candidate. Here we consider a minimal UED model with one extra dimension $X=S^1/Z_2$ with a compactification radius $R$. Indirect constraints on $R$ from electroweak precision tests and the dark matter relic density favour a mass scale of the first Kaluza-Klein modes of ${\cal O} (1\!\tev)$. UED models can thus be directly probed at the LHC, either through signatures with missing transverse momentum, or through searches for resonances near the TeV scale. Since the mass spectrum of the Kaluza-Klein resonances is rather compressed,  
searches based on missing transverse momentum signatures suffer in general from low detection efficiencies. Here we investigate the search for the production of second Kaluza-Klein modes through resonant decays into Standard Model particles. Such signatures are of particular interest as they would allow for a clear distinction between UED and minimal supersymmetric models. Specifically, we consider the signature $pp\rightarrow A^{(2)}/Z^{(2)}+X\rightarrow l \bar{l}+X'$, where second Kaluza-Klein electroweak gauge bosons decay into electron or muon pairs, and constrain the model by comparing the predicted minimal UED signal rates to the results of a recent CMS search for dilepton resonances in the 8\,TeV run with 20.6\,fb$^{-1}$ integrated luminosity for dimuons and 19.6\,fb$^{-1}$ for dielectrons.  In the absence of any signal we place limits on the compactification scale  $R^{-1} \gtrsim 715$\,GeV and on the masses of the second Kaluza-Klein particles  $M_{{\rm KK}^{(2)}} \gtrsim 1.4$\,TeV. 

\bigskip

\noindent
{\bf Note added:} The UED constraints in this article are based on the CMS search for dilepton resonances \cite{cmsdilepton}. After submission of this article, ATLAS made its search for high-mass dilepton resonances public \cite{ATLAS:2013jma}. We  adopted and re-ran our analysis using the ATLAS constraints, and found marginally weaker constraints on minimal UED as compared to the constraints presented here.

\bigskip

\noindent
\emph{Acknowledgements:} \\
The authors thank Stefan Schmitz for helpful discussion on the CMS dilepton search. 
This work has been supported in part by the Helmholtz Alliance ``Physics at the Terascale'' and the DFG SFB/TR9 ``Computational Particle Physics''. TF was supported by the National Research Foundation of Korea (NRF) grant funded by the Korea government (MEST) N01120547. He also thanks CERN for hospitality where part of this work has been done.

\end{document}